# Stable, predictable and training-free operation of superconducting Bi-2212 Rutherford cable racetrack coils at the very high wire current density of more than 1000 A/mm$^2$

Tengming Shen[1,*], Ernesto Bosque[2], Daniel Davis[1,2], Jianyi Jiang[2], Marvis White[3], Kai Zhang[1], Hugh Higley[1], Marcos Turqueti[1], Yibing Huang[4], Hanping Miao[4], Ulf Trociewitz[2], Eric Hellstrom[2], Jeffrey Parrell[4], Andrew Hunt[3], Stephen Gourlay[1], Soren Prestemon[1], David Larbalestier[2]

[1] Lawrence Berkeley National Laboratory, Berkeley, CA 94720, USA

[2] National High Magnetic Field Laboratory, Florida State University, Tallahassee FL 32310, USA

[3] nGimat LLC, Lexington, KY 40551, USA

[4] Bruker OST LLC, Carteret, NJ 07008, USA

*Email: tshen@lbl.gov

Abstract

High-temperature superconductors (HTS) could enable high-field magnets much stronger than is possible with Nb-Ti and Nb$_3$Sn, but two key limiting factors have so far been the difficulty of achieving high critical current density in long-length conductors, especially in high-current cables, and the danger of quenches out of the superconducting into the normal state. Most HTS magnets made so far have been made out of REBCO coated conductor, which is now available in lengths of 200-300 m. Here we demonstrate stable, reliable and training-quench-free performance of Bi-2212 racetrack coils wound with a 17-strand Rutherford cable fabricated from wires made with nanospray Bi-2212 powder. These multifilament wires are now being delivered in single lengths of more than 1 km with a new record whole-wire critical current density up to 950 A/mm$^2$ at 30 T at 4.2 K. These coils carried up to 8.6 kA while generating a peak field of 3.5 T at 4.2 K, at a wire current density of 1020 A/mm$^2$. Quite different from the unpredictable training performance of Nb-Ti and Nb$_3$Sn magnets, these Bi-2212 magnets showed no training quenches and entered the flux flow state in a stable manner before thermal runaway and quench occurred. Also quite different from Nb-Ti, Nb$_3$Sn, and REBCO magnets for which localized thermal runaways occur at unpredictable locations, the quenches of Bi-2212 magnets consistently occurred in the high field regions over a conductor length greater than one meter. These characteristics make quench detection rather simple, enabling safe protection, and suggest a new paradigm of constructing quench-predictable superconducting magnets from Bi-2212, which is, like Nb-Ti and Nb$_3$Sn, isotropic, round, multifilament, uniform over km lengths and suitable for Rutherford cable use but, unlike them, much more tolerant of the energy disturbances that often lead Nb-based superconducting magnets to premature quench and long training cycles.

High field superconducting magnets are used in particle colliders[1], fusion energy devices[2], magnetic resonance imaging (MRI), ion beam cancer therapy[3], as well as thousands of nuclear magnetic resonance (NMR) and general laboratory magnets.  So far, virtually all superconducting magnets have been made from two Nb-based low temperature superconductors (Nb-Ti with superconducting transition temperature $T_c$ of 9.2 K and $Nb_3Sn$ with $T_c$ of 18.3 K).  The 8.33 T Nb-Ti accelerator dipole magnets of the large hadron collider (LHC) at CERN enabled the discovery of the Higgs Boson and the ongoing search for physics beyond the standard model of high energy physics.  $Nb_3Sn$ magnets are key to the International Thermonuclear Experimental Reactor (ITER) Tokamak and to the high-luminosity upgrade of the LHC[4] that aims to increase the luminosity by a factor of 5-10.

Nb-Ti and $Nb_3Sn$ magnets predominately work at 1.8-5 K, generating magnetic fields seldom greater than 20 T.  The superconducting magnet community has long sought to broaden the superconducting application space to higher fields at low temperatures (4.2-20 K) or higher temperatures (20-77 K) using high-temperature superconductors (HTS)[5].  Cuprate superconductors, discovered in the 1980s, and iron-based superconductors, discovered about 10 years ago, both have upper critical magnetic fields ($H_{c2}$) exceeding 50 T at 4.2 K[6-8], much greater than that of Nb-Ti (~14 T at 1.8 K) and $Nb_3Sn$ (~26-27 T at 1.8 K), making them potentially useful for making much stronger high-field magnets[9], like the 30 T superconducting solenoids needed for 1.3 GHz NMR spectrometers[10] and 20 T accelerator dipoles for a potential high-energy upgrade of the LHC[11].  Significant progress has been made, particularly in solenoids made of $REBa_2Cu_3O_x$ (REBCO), culminating in a 32 T user solenoid recently constructed at the National High Magnetic Field Laboratory (NHMFL)[9,12].  After three decades of arduous conductor development, high-temperature cuprate superconducting conductors, including $Bi_2Sr_2CaCu_2O_x$ (Bi-2212)[13], $(Bi,Pb)_2Sr_2Ca_2Cu_3O_x$ (Bi-2223)[14], and REBCO[9], are being commercially produced in practical forms of metal/superconductor composite conductors in lengths suitable for making magnets, though long-length REBCO tapes often have localized processing defects [15].  These materials can deliver high critical current density in strong magnetic fields at 4.2 K or 1.8 K above 23 T, where the $J_c$ of $Nb_3Sn$ wires ceases to be useful.

All superconducting magnets are susceptible to quench[16,17] when local regions lose superconductivity, producing hot spots with rapidly rising temperature that may, without suitable protection, lead to local degradation or burnout.  An important drawback of Nb-Ti and $Nb_3Sn$ superconducting magnets is their low thermal stability against quench.  Tiny, transient point disturbances as small as $10^{-6}$ J from conductor motion are very common due to the large thermal and electromagnetic stresses and they can cause localized temperature rises sufficiently large to initiate quench of the whole magnet. The small thermal margin of Nb-Ti and Nb$_3$Sn and their low heat capacity below $T_c$ give them little protection against such energy inputs. To overcome such instabilities, the main-ring LHC Nb-Ti dipole and quadrupole magnets are cooled by 1.8 K superfluid helium that penetrates into the windings, where it acts both as a powerful coolant due to its high thermal conductivity and a thermal stabilizer due to the very high specific heat of superfluid He which is much higher than that of the oxygen-free-high-conductivity copper (OFHC) that thermally stabilizes and protects the Nb-Ti cable during quench. The quench problem is also a costly and still unsolved issue for the Nb$_3$Sn magnets being considered for future high-energy proton colliders and a high-energy upgrade of the LHC, in spite of its $T_c$ being twice that of Nb-Ti (18 versus 9 K). The reason is that Nb$_3$Sn windings cannot be permeated with superfluid helium because they must be epoxy-impregnated to protect them against the large crushing forces in the magnets. Epoxy-impregnated magnets generally experience epoxy cracking and interfacial shearing and debonding and the poor winding thermal conductivity and energy disturbances produced by such cracking often results in long quench training, currently as many as 20-30 quenches for the HL-LHC Nb$_3$Sn quadrupole magnets[18]. Such training results in significant helium loss and high labor costs, significant drawbacks in considering future high-energy proton-proton colliders that need thousands of such magnets. For other smaller scale, one-of-a-kind physics experiments, this problem can also be severe. An example is MICE, the Muon Ionization Cooling Experiment, built with a cryocooler-cooled, epoxy-impregnated Nb-Ti conductor. Both MICE solenoids underwent a long training despite using a highly stabilized wire with a high (3:1) Cu to superconductor ratio. Due to its high stored magnetic energy, each quench of the Nb-Ti spectrometer solenoid evaporated 800 L of liquid helium. Even worse, the MICE spectrometer solenoids had a poor training memory and needed retraining after thermal cycling[19]. The cost of training these magnets, despite the low cost of Nb-Ti wires, was a big financial distress.

Due to their much higher $T_c$ and $H_{c2}$, HTS magnets have much higher enthalpy stability margin against quench, making quench less likely though this remains to be proved experimentally, especially for accelerator magnets[20]. Despite frequent Nb-Ti and Nb$_3$Sn magnet quenches, protection during quench is generally not an issue because quench propagation velocities are rapid and the stored magnetic energies can be safely dissipated over large volumes of the coil or transferred to an external protection circuit. By such methods, even magnets with wire engineering (i.e. whole wire) current density $J_E$ of 1000 A/mm$^2$ have rarely been damaged. However, the paradox for HTS magnets is that, when quench does occur, it is a much bigger threat than in LTS magnets because normal zones are smaller, quench velocities much smaller, with the consequence of much higher energy density in much smaller normal zones. Several REBCO and Bi-2223 magnet systems have been degraded during quench[21-27], likely when small normal zones generated too little voltage for warning, making active quench protection unable to prevent local overheating as magnetic energy turned into very localized heating. A key challenge of HTS magnets is thus the timely detection of small hot spots that grow only at several cm/s, rather than the two orders of magnitude larger rate in Nb-Ti and $Nb_3Sn$ magnets.

All of these concerns are very much reduced in the Bi-2212 magnets described here, even though they were quenched at a new record, high wire current density $J_E$ of up to 1000 A/mm$^2$. Greatly assisting this performance is the use of a rather uniform, isotropic, multifilamentary HTS Bi-2212 round wire Rutherford cable[13]. We experimentally demonstrate quench-training-free operation of epoxy-impregnated racetrack coils at currents above 8 kA, values that exceed the benchmark current and $J_E$ values needed for demanding high-energy physics collider magnets. We further suggest a new paradigm of constructing quench-predictable superconducting magnets from Bi-2212, and explain how and why quench can be managed with Bi-2212.

**Methods**

*Wire and cable design and fabrication*

A 440 m long, 0.8 mm diameter Bi-2212 round wire was fabricated by Bruker OST LLC by their standard powder-in-tube technique using a novel precursor Bi-2212 powder made at nGimat LLC by a nano-spray combustion technology. This new powder appears to eclipse the earlier industrial benchmark Bi-2212 powder made by the melt-casting approach at Nexans[28]. The wire has an architecture of 55 x 18 (18 bundles of 55 filaments). The Bi-2212 filling factor of the as-drawn wire is ~25% and porosity occupies roughly 30% of the filament cross-section as delivered. The matrix Ag surrounding the filaments has a very high electrical conductivity as judged by a resistance ratio >100 and very high thermal conductivity[29,30]. 17-strand Rutherford cables were made at LBNL with a width of 7.8 mm and a thickness of 1.44 mm. The cable insulation was a braided mullite sleeve with a wall thickness of ~150 μm.

*Coil design, fabrication, and test*

Two 2-layer, 6-turns per layer subscale racetrack coils (RC5 and RC6) without any internal joints using 8 m long Rutherford cables with 140 m of 0.8 mm wire in each cable were wound on an Inconel 600 pole island. Each coil was assembled inside an Inconel 600 structure of side bars and top and bottom plates and then heat treated by an overpressure processing heat treatment (OPHT) technology at the NHMFL with a gas pressure of 50 bar in flowing Ar/$O_2$ (oxygen partial pressure $P_{O2}$ = 1 bar). OPHT removes most of the starting porosity in the wire, increasing the filament density to >95%. The coil and its reaction structure weighs ~8 kg and measures 37 cm x 12 cm x 3.1 cm. After reaction, the coil was instrumented with voltage taps on each turn, and impregnated with epoxy resin (RC5 used the rather brittle CTD 101K epoxy adopted for High-Luminosity LHC (HL-LHC) $Nb_3Sn$ magnets, while the more fracture-resistant NHMFL "mix-61" epoxy, developed for the large-bore 900 MHz NMR magnet was used for RC6).

The coils were tested inside their Inconel reaction structure. They were powered with a 20 V, 24 kA DC power supply, while the terminal voltages were monitored using a fast, FPGA-based quench detection system. Upon detecting a quench, the FPGA board sent commands to open an SCR electrical switch that inserted a room temperature dump resistor (20 m$\Omega$) across the coils, forcing the magnet current to decay to zero within <10 ms. The magnet voltage was recorded using a 18-bit ADC system with a programmable isolation amplifier and a flexible, software-controlled measurement range that could be set from $\pm$0.1 mV to $\pm$5 V at both 10 Hz and 1 kHz.

**Results**

***The high critical current density of these Bi-2212 wires***

The 4.2 K wire current density $J_E$ of the optimally processed strand used in this study is shown in Figure 1 [28]. $J_E$ is greater than that of the HL-LHC RRP (Rod Restack Process) $Nb_3Sn$ strand above 11 T and it also has a much less field-sensitive characteristic, achieving 1365 A/mm$^2$ at 15 T, twice the target desired by the Future Circular Collider (FCC) for $Nb_3Sn$ strands[31], and 1000 A/mm$^2$ at 27 T, 65% better than the previous record Bi-2212 performance[32].

***Quench performance of the coils***

Ramping up the current of any superconducting magnet eventually leads to a quench, and this was no different for RC5 and RC6. The coil voltages of RC5 and RC6 shown in Figure 2 exhibit characteristic quench behavior, with $V_{13}$ (the whole coil terminal voltage) and $V_{12}$ (one half of the coil) going positive due to one half coil entering the dissipative state first, while $V_{23}$ (the second half the coil) trends negative due to its inductive response to the growing normal zone in coil 1. Despite the small coil inductance (~35 $\mu$H), the voltage noise is on the order of mV, making it difficult to set the quench detection voltage to be less than 10 mV.

The quench current of RC5 and RC6 actually increased slightly on raising d$I$/d$t$ from 30 A/s to 200 A/s (Figure 3) and they exhibited no quench training at all. $I_q$ remained nearly unchanged (average current = 8604 A, standard deviation = 4.1 A for RC6) during consecutive quenches and after thermal cycling to room temperature and back to 4.2 K (Figure 4) ($I_q$ did drop about ~0.5% after the thermal cycling to 4.2 K but remained unchanged throughout the second test. No second thermal cycling test was performed).

To examine the nature of the RC5 and RC6 quenches, the magnets were powered up using a staircase scheme, during which the coil current was periodically held constant so as to zero inductive voltages, minimize noise and allow the maximum insight into dissipation within the coils. The terminal and turn-to-turn voltages of RC5 (Figure 5) show that several turns had started to dissipate before the terminal voltage $V_{13}$ exceeded 0.1 V, which indicates that quench was triggered by a significant length of cable entering the dissipative state, rather than being triggered by point disturbances as is generally the case for $Nb_3Sn$ magnets. The signals taken during stable-current portions of the staircase ramp show that resistive voltage signals became visible at 7000 A well below the >8000 A of $I_q$ and that they steadily increased with increasing current. From $t$ = 447 s to $t$ = 507 s, the current was held at 7925 A and the resistive voltage of the ramp turn (a 14 cm long section that transitions between the two coil layers in the peak field region. See Figure 5c) was 3.6 µV, generating a steady joule heating of 28.5 mW. The total heat input during this hold from 447 to 507 s was ~1.71 J, which, though significant, did not cause the coil to quench. When the coil current was raised and held at 8130 A, the resistive voltage of the ramp turn was 9.4 µV and the joule heating 76.4 mW, which caused a thermal runaway in less than 4 seconds. Such a high thermal stability is in strong contrast to the instability that characterizes Nb-Ti and $Nb_3Sn$ magnets, for which a disturbance as small as 1 µJ is sufficient to cause the coil to quench.

### *The global superconducting-normal transition: Coil E-I curve and $I_c$*

Figure 6 plots the *E-I* curve of RC5 and RC6, where the electric field *E* is derived from the resistive signals during the current-hold parts of the staircase current ramps when the inductive ramping signals were absent. Figure 6 suggests that the resistive signals were driven by usual power law resistive transition losses ($V \sim I^n$, where *n* is 20-30) during the smooth transition of the Bi-2212 cable from the superconducting to the normal state. The critical current $I_c$, defined at an electric field criterion of 0.1 µV/cm, was 7550 A for RC5 and 7750 A for RC6. The *n*-values determined by the power law fitting the *E-I* curve were 22 for RC5 and 24 for RC6, values very similar to those obtained in short single wire samples. This speaks to the excellent uniformity and current sharing of the Rutherford cables used in these two magnets.

**Discussion**

The baseline magnet technology for future high energy proton colliders is the 12-16 T wind-and-react, epoxy-impregnated Nb$_3$Sn magnets with lengthy and costly training. An important and novel outcome of this work is experimental demonstration of quench-training-free operations of epoxy-impregnated Bi-2212 magnets at the very high current density of 1000 A/mm$^2$. This performance provided a proof-of-principle that Bi-2212 should be considered for the 16-20 T accelerator magnet technology for future colliders. It is very important that neither RC5 nor RC6 quenched as a result of transient, tiny, and localized disturbances that normally give early and unpredictable quenches below $I_c$ for Nb-Ti and Nb$_3$Sn magnets. Stable operation in the dissipative state was observed in RC5 and RC6, three other impregnated coils RC1 (wax), RC2 (brittle CTD-101K epoxy), and RC3 (NHMFL mix-61 tough epoxy)[33], and coils made from Bi-2212 tapes[34]. Another sign of the great stability of RC5 and RC6 is the insensitivity of their quench current to ramp rate. Indeed, these Bi-2212 magnets quenched due to continuous joule heating when the cable current approached the natural transition of the cable from the fully superconducting to the partially dissipative state of the critical current transition.

A vital concern for any HTS magnet is the need to detect quench so that active protection can be triggered. The greatest danger is localized thermal runaway at unexpected locations. Long quench training of Nb$_3$Sn magnets is a cumbersome and costly problem but it seldom risks coil loss. In contrast, failing to detect a quench in a much more stable HTS magnet puts the HTS magnet in great danger because the dissipative zones propagate only slowly and overheat quickly. At the very high wire current density of 1000 A/mm$^2$ safely demonstrated in these Bi-2212 racetrack magnets, hot spot temperatures can rise at hundreds of K/s, thus making rapid (<10 ms) detection and rapid switching of the dump resistor vital to a safe quench without magnet burnout. The state-of-the-art quench detection for Nb-Ti and Nb$_3$Sn magnets relies on fast (>10 k-samples/s) but low precision (mV resolution) voltage measurements to define a typical threshold in the range of 100 mV to several volts. Due to the fast quench propagation velocities (typically 10 m/s for Nb-Ti and Nb$_3$Sn magnets), 100 mV develops across growing normal zones in a negligible 0.1 ms. By contrast, at the typical 1 cm/s quench velocities of HTS conductors, resistive voltages across a localized hot spot increase at less than 0.1 mV within 0.1 ms. Most HTS magnets made so far have been made out of REBCO coated conductor. Important concerns with REBCO conductors, despite their high stability, are that they contain localized defects either pre-existing [15] or created during coil winding or during magnet operation [22], that their critical current is anisotropic in magnetic field, and that this anisotropy depends on flux pinning and therefore REBCO processing conditions; these uncertainties mean that quench is often localized, unpredictable, unexpected, and catastrophic [21-23,27]. This is illustrated by a single-layer coil wound from a 2 m long REBCO single tape (2 mm wide, 100 μm in thickness with 40 μm Cu) with a known defected 2 cm section ($I_c$ = 48 A versus ~55 A for other sections at 77 K). This coil went into a thermal run-away at 60 K at ~200 A, with the last recorded terminal voltage of 1.2 mV (DAQ rate at 10 samples/s) and the voltage across its defective section of 0.4 mV, resulting in unexpected burnout in another coil section. In spite of these dangers, the default quench detection method for REBCO magnets is still a voltage measurement in the mV range with a control limit of several or tens of mV[21-25]. It is thus no surprise that many REBCO magnets have been damaged during quench.

An important advantage of Bi-2212 coils revealed by this work is that their quench is global and essentially predictable. Several meters of the Bi-2212 cable in the high field region developed stable voltages (tens to hundreds of μV) well before thermal runaway and they then enter thermal runaway nearly simultaneously, as observed in RC1, RC2, and RC3[33] and a single-strand (140 m long), epoxy impregnated Bi-2212 solenoid tested in a background field of 14 T[35]. The predictable quench behavior of these Bi-2212 coils is due to the uniqueness of Bi-2212 as a round, isotropic multifilament HTS wire, now available in km-lengths with excellent current sharing in classic Bi-2212 Rutherford cables (evidenced by a *n*-value of greater than 20 (Figure 6)). By contrast, REBCO strands are monofilaments with many localized defects[15] and do not carry current uniformly due to their large critical current anisotropy and large screening current effects.

A key understanding revealed in this letter is that quench locations of Bi-2212 magnets are predictable. Unlike Nb-Ti and Nb$_3$Sn magnets that do not always quench in their high field regions due to their vulnerability to random point disturbances, and unlike REBCO magnets that do not always quench in the high field regions due to localized defects and their large temperature-dependent anisotropy, a stable transition is observable before quench for Bi-2212 magnets and they consistently enter into thermal runaway in their high field regions, where quench is expected. An important technological consequence is that standalone Bi-2212 magnets may be made quench-free, since their operation margins can be understood quantitatively by measuring the rise of resistive voltage in the high field region with a staircase current ramp with an accuracy of 0.1 μV, which is an improvement of orders of magnitude over the sensitivity of quench detection used by Nb-based magnets.

Our work therefore shows why and how quenches can be detected for Bi-2212 magnets. Its >20 T high-field magnet applications are most likely going to be hybrid ones with a background magnetic field provided by Nb-Ti and $Nb_3Sn$ outsert magnets. Through magnetic coupling quench of Nb-Ti and $Nb_3Sn$ magnets would trigger the HTS magnet to quench. In the case of the Nb-Ti/$Nb_3$Sn/REBCO 32 T magnet [12], the inner 17 T, 34 mm bore, ~100 kg REBCO magnet is powered with an independent power supply circuit and has to be protected with a total heater energy of 160 kJ, supplied by a high voltage battery system, in comparison to 300 J for active protection of the outsert 1400 kg, 15 T, 250 mm bore Nb-Ti/$Nb_3$Sn magnet [9,36,37]. Bi-2212 presents a feasible path to developing commercial, more user-friendly 25-30 T solenoid systems than the 32 T magnet. Being a round wire with a flexible diameter and current-carrying capability, Bi-2212 inserts can be placed in electrical series with the background Nb-Ti and $Nb_3$Sn magnets, so that the quench protection of such system can be much simpler.

We finally emphasize that the stable and safe quench performance described here was obtained on wires with the very high wire critical current density of 1000 A/mm$^2$, current densities well above those required for efficient dipole magnet design. We emphasize that even higher wire current densities of 1800 A/mm$^2$ at 5 T and 4.2 K and 1000 A/mm$^2$ at 27 T and 4.2 K are available from short-samples of these wires [13,32]. As usual, there is a gap between magnet and best short sample performance, but it is important to note that recent Bi-2212 conductor improvement has passed into the magnets too. The quench current of RC6 is 8.6 kA, whereas the best performance coil RC2 made with previous generation wires [13] using Nexans granulate powder had $I_q$ of 5.7 kA [33]. Bi-2212 cables have leapt beyond the benchmark wire $J_E$ needed by high-energy physics particle collider magnets, typically made of Rutherford cables with an average cable current density 400-750 A/mm$^2$ and a wire $J_E$ of 600-1000 A/mm$^2$. Bi-2212 wires now enable such current densities for 20-T class dipoles. The breakthrough was made possible by better powders made by a combustion chemical vapor condensation[38,39,40], which eclipse the previous melt-cast[41], aerosol spray pyrolysis and liquid co-precipitation powders [13,33]. Now it appears that highly homogeneous, 100-500 nm powders with excellent composition control (+/-2% for Ca and +/-1% for others), a low impurity level, and low risk of powder agglomeration can be made. Such powders enable uniform drawing of wires with fine superconductor filaments (10-20 μm in diameter) by the powder-in-tube method.

**Conclusions**

We have demonstrated that the exceptionally high wire current density of 1000 A/mm$^2$ can be reached in an HTS Bi-2212 magnet and that the onset of thermal runaway is nonlocalized, enabling safe protection against quench. Quenches consistently occurred in the high field regions and exhibited a highly uniform resistive transition of the cable that permits detection of the resistive transition voltages well before quench. This high wire current density was made possible by a new nanospray combustion powder technology. The feasible quench detection and possible quench-free operation for standalone Bi-2212 magnets enables magnet use of the unique characteristics so far only found in Bi-2212, a combination of high stability that comes with high-$T_c$ cuprate superconductors, isotropic behavior due to the round, multifilamentary wire architecture, and the ability to produce long wires and cables with high uniformity and current sharing. With these capabilities, Bi-2212 has taken a huge step towards practical applications such as user-friendly, 25-30 T class solenoid research magnets [42], 20-T class, canted cosine theta geometry[43,44], accelerator magnets useful for future high-energy colliders, including a high-energy LHC upgrade, and >1.3 GHz NMR magnets due to Bi-2212 being the only isotropic, multifilamentary HTS round wire that permits high magnetic field quality.

**Acknowledgements**

Work at LBNL was supported by the Director, Office of Science of the U.S. Department of Energy (DOE) under Contract No. DE-AC02-05CH11231. Work at the NHMFL was supported by the US DOE Office of High Energy Physics (OHEP) under grant number DE-SC0010421, by the National Institutes of Health under Award Number R21GM111302, and by the NHMFL, which was supported by the National Science Foundation under Award Numbers DMR-1157490 and DMR-1644779, and by the State of Florida. Work at Bruker OST and nGimat was supported by the U.S. DOE OHEP through a SBIR award DE-SC0009705. This work was amplified by the U.S. Magnet Development Program (MDP). T.S. acknowledges support from the U.S. DOE Early Career Research Program, K.Z. acknowledges support from the China Scholarship Council, and D.D. acknowledges support from the U.S. DOE Office of Science Graduate Student Research Program.

**Author contributions**

**Additional Information**

The author(s) declare no competing interests.

**List of Figures**

Figure 1: $J_E(B)$ of an optimally processed sample of the strand used in this study in comparison to that of a Bi-2212 with the previous record performance, the LHC Nb-Ti strand, and the HL-LHC RRP $Nb_3Sn$ strand.

Figure 2: Coil voltages (RC5) during a linear current ramp (see inset in b) that ended with a quench. **a** Voltage tap map. **b** Coil voltages $V_{13}$ (whole coil) and $V_{12}$ and $V_{23}$ (individual layers).

Figure 3: Ramp rate dependence of the quench current $I_q$ of RC5 and RC6 during linear current ramps. Inset shows a 3D display of the contours of the surface magnetic flux density generated by RC6 at 8600 A.

Figure 4: $I_q$ of RC6 for consecutive quenches before and after thermal cycling to room temperature and back to 4.2 K.

Figure 5: Voltage development of RC5 for staircase ramps of the magnet current that ended with thermal runaway and energy extraction by switching in a dump resistor. The current ramp scheme contains current holding steps during which coil inductive signals die away and noise is much reduced **b**. The coil and turn-to-turn voltages are shown in **b** and **d**, respectively. The ramp turn voltage is highlighted in **c**. The ramp turn is a 14 cm long section that transitions between the two coil layers in the peak field region. In **d**, L1-T1 means the turn #1 of the coil layer #1 (other turns follow the same naming method.) and it is the outermost turn in the low field region. The voltage tap length of turns decreases from 62 cm for L1-T1 and L2-T1 gradually to 52 cm for L1-T6 and L2-T6.

Figure 6: The *E-I* transition of RC5 and RC6 derived from tests with staircase powering schemes for the ramp turns.

Figure 1:

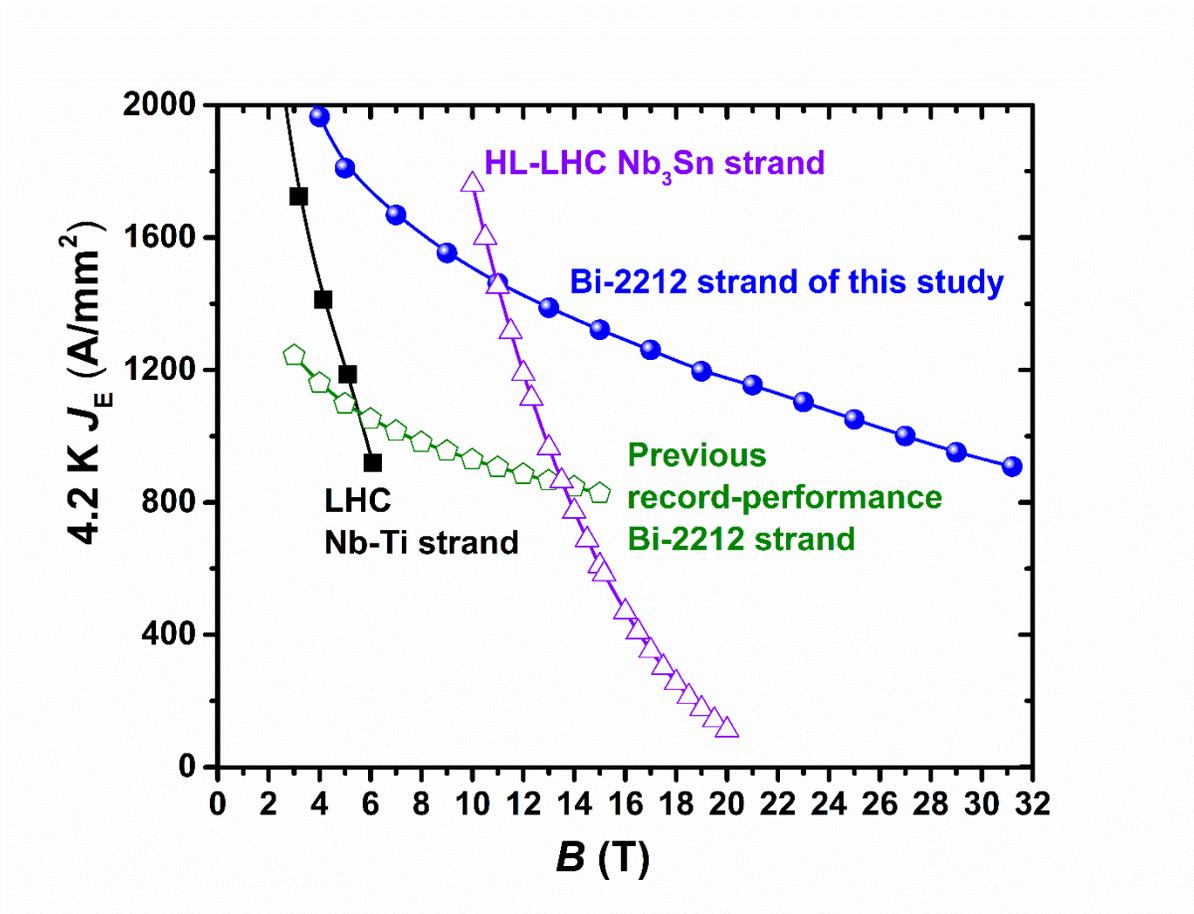

Figure 2:

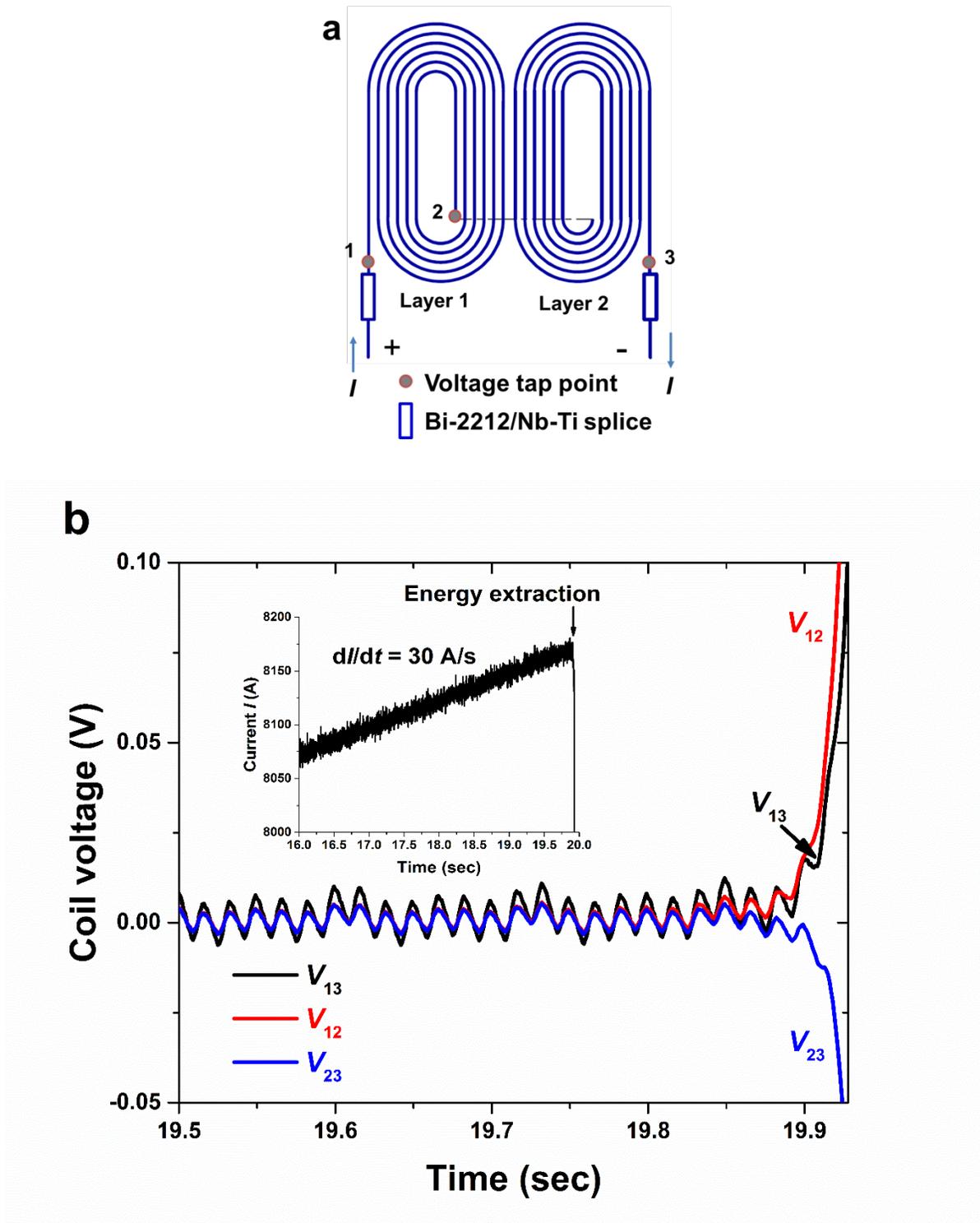

Figure 3:

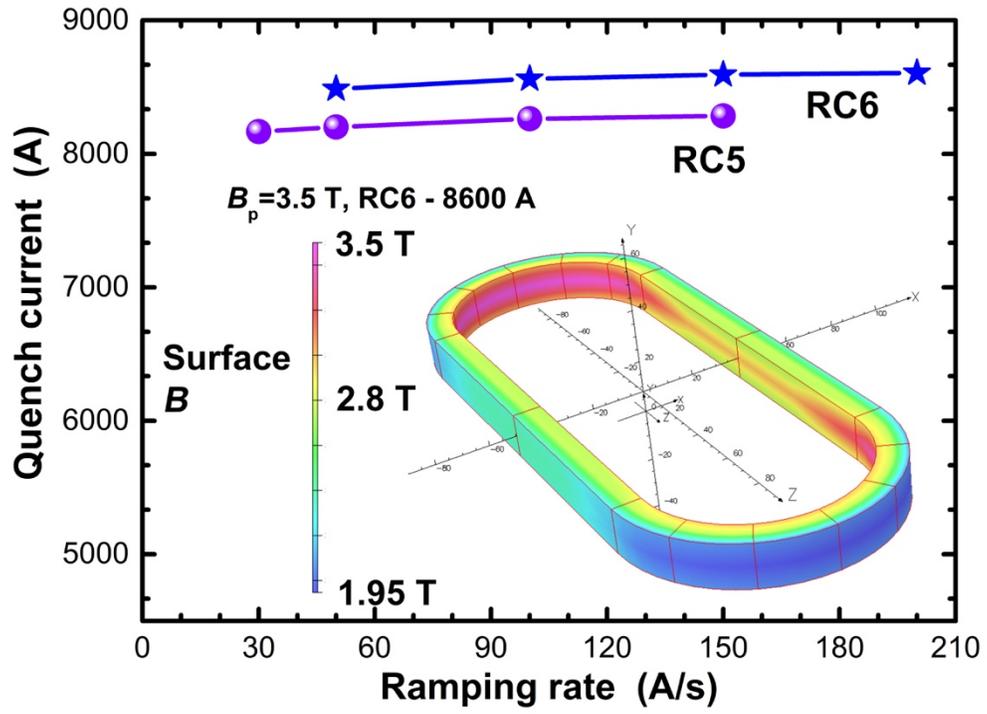

Figure 4:

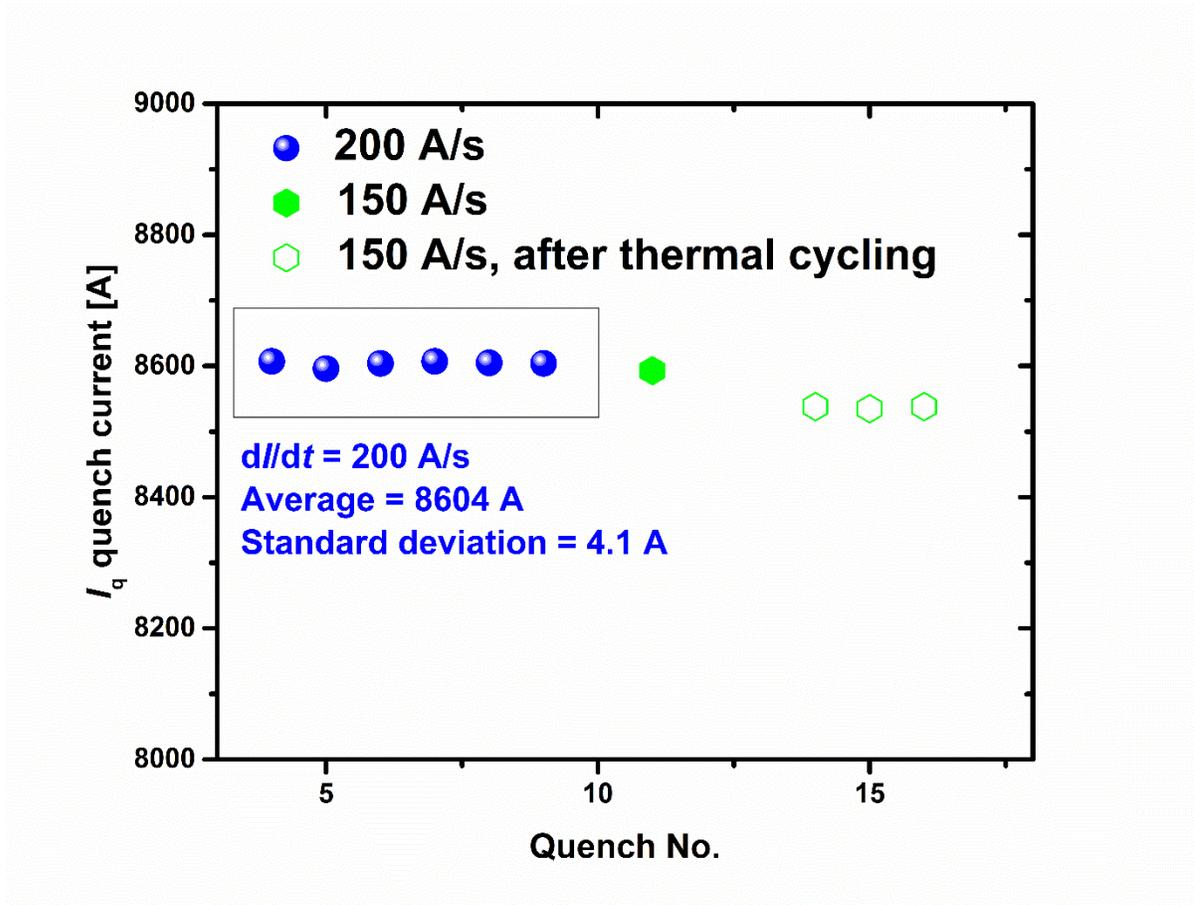

Figure 5:

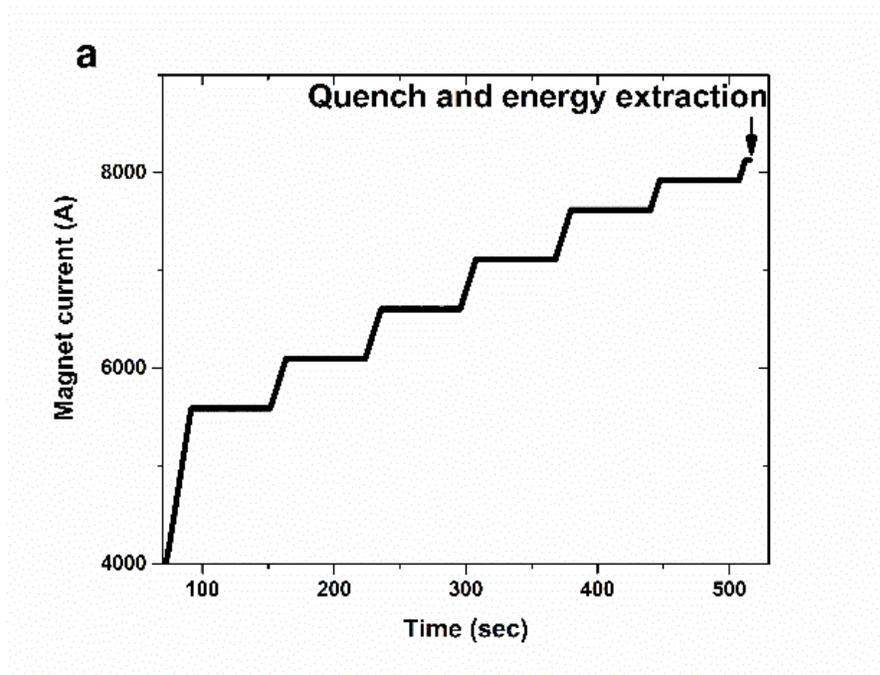

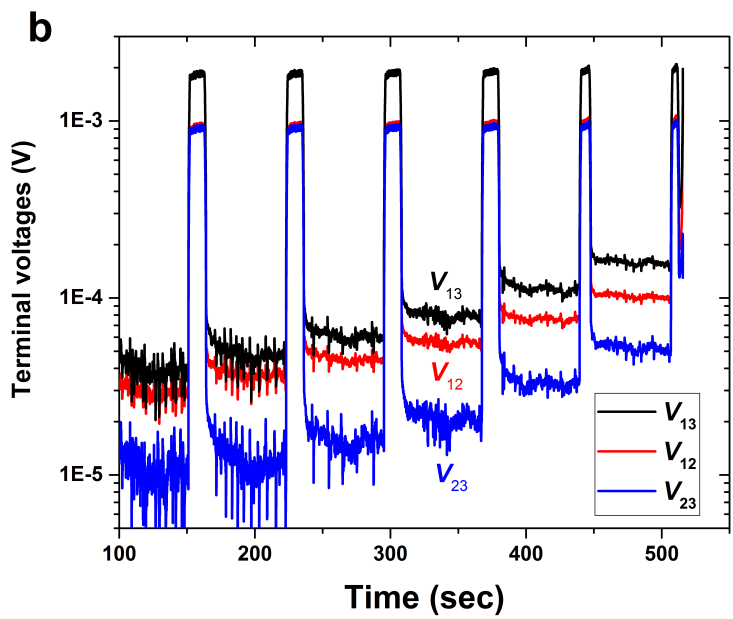

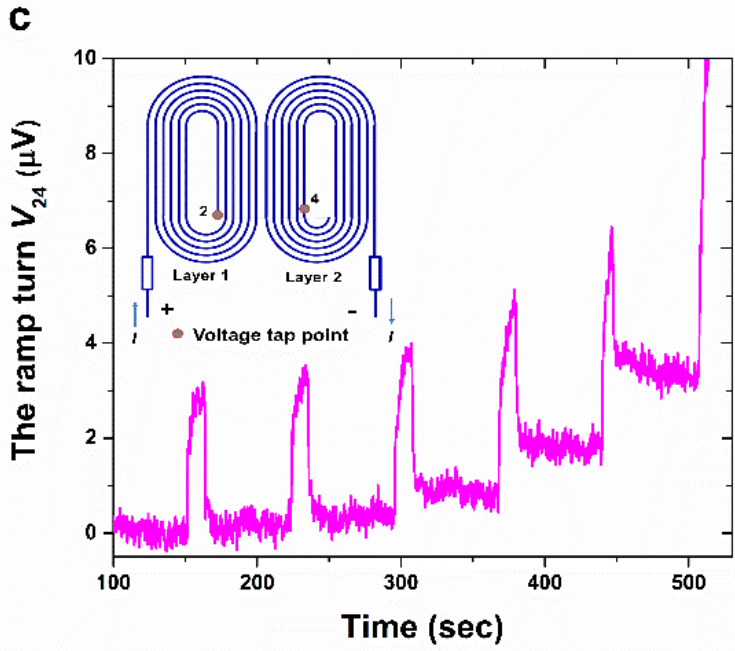

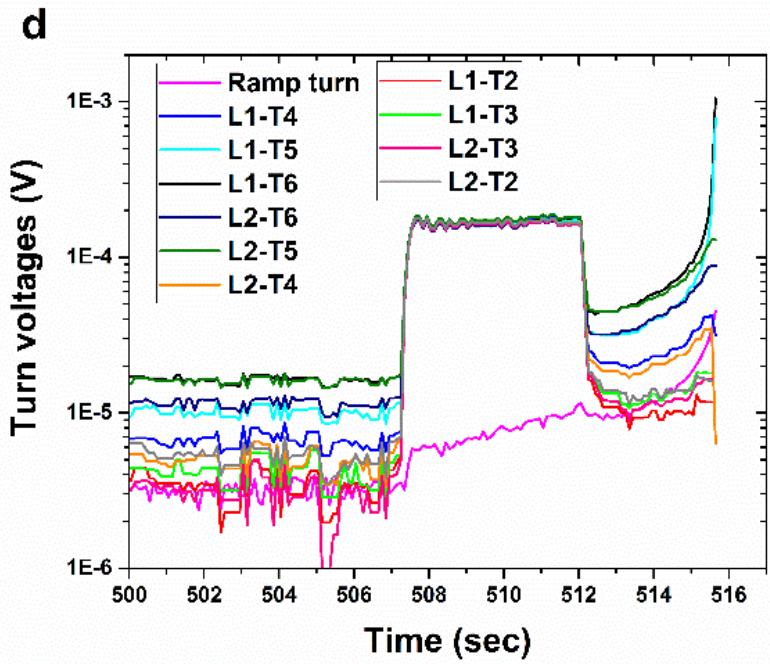

Figure 6:

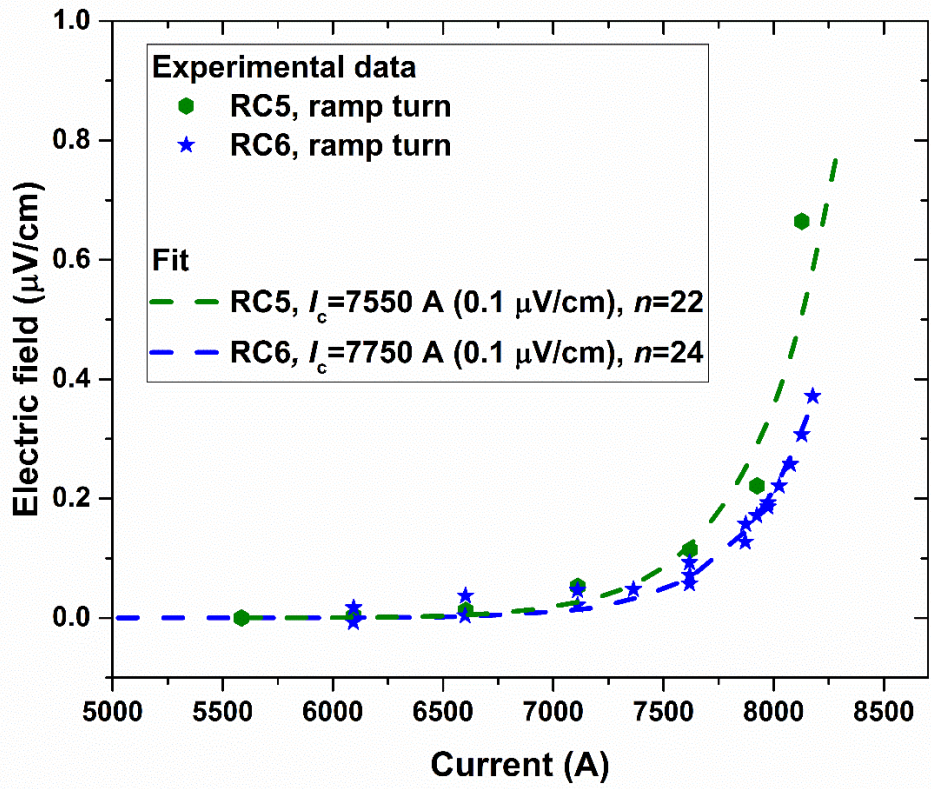